\documentclass[aps,twocolumn,superscriptaddress,floatfix]{revtex4-1}
\usepackage{bbm} 
\usepackage{amsmath} 
\usepackage{amssymb} 
\usepackage{physics} 
\usepackage[bookmarks=true]{hyperref} 
\usepackage[english]{babel} 
\usepackage[normalem]{ulem} 
\usepackage{xfrac}
\usepackage{graphicx}
\usepackage[]{xcolor} 

\allowdisplaybreaks

\begin{document}

\title{Geometric Phase of a Transmon in a Dissipative Quantum Circuit}

\author{Ludmila Viotti} 
\affiliation{ The Abdus Salam International Center for Theoretical Physics, Strada Costiera 11, 34151 Trieste, Italy} 
\author{Fernando C. Lombardo}
\affiliation{Departamento de Física, Facultad de Ciencias Exactas y Naturales, Universidad de Buenos Aires, Buenos Aires, Argentina}
\affiliation{ Instituto de Física de Buenos Aires (IFIBA), CONICET, Universidad de Buenos Aires, Argentina}
\author{Paula I. Villar}
\affiliation{Departamento de Física, Facultad de Ciencias Exactas y Naturales, Universidad de Buenos Aires, Buenos Aires, Argentina}
\affiliation{ Instituto de Física de Buenos Aires (IFIBA), CONICET, Universidad de Buenos Aires, Argentina}

\begin{abstract} 
Superconducting circuits reveal themselves as promising physical devices with multiple uses. Within those uses, the fundamental concept of the geometric phase accumulated by the state of a system shows up recurrently, as, for example, in the construction of geometric gates. Given this framework, we study the geometric phases acquired by a paradigmatic setup: a transmon coupled to a superconductor resonating cavity. We do so both for the case in which the evolution is unitary and when it is subjected to dissipative effects. These models offer a comprehensive quantum description of an anharmonic system interacting with a single mode of the electromagnetic field within a perfect or dissipative cavity, respectively. In the dissipative model, the non-unitary effects arise from dephasing, relaxation, and decay of the transmon coupled to its environment. Our approach enables a comparison of the geometric phases obtained in these models, leading to a thorough understanding of the corrections introduced by the presence of the environment.
\end{abstract}

\maketitle


\section{Introduction}\label{sec:intro}

Significant advancements in coherent superconducting circuits have enabled the development of a diverse range of qubit designs, encompassing the classic flux~\cite{1E,2E,3E,4E}, phase~\cite{5E,6E,7E,8E}, and charge~\cite{9E,10E,11E} qubits, as well as more contemporary designs like transmon~\cite{12E} and fluxonium~\cite{13E} circuits. The mathematical representation and quantum dynamics of these qubits, when transversely coupled to resonators, are described by circuit quantum electrodynamics (cQED)~\cite{Blais,review_girvin2014circuit, review_makhlin2001quantum}. Although these qubits are designed to behave as a two-level system within a superconducting circuit, they inherently possess multiple additional energy levels that can influence their interactions with other components. Focusing on the transmon qubit, one of its main advantages is its longer coherence timescale when compared to other circuits, which is essential for performing quantum computations and quantum error correction operations~\cite{marques2023all, battistel2021hardware, kim2023design}. This increased coherence can be attributed to the non-harmonic energy level structure, which helps suppress certain types of \mbox{decohering effects~\cite{chen2023transmon}.}

Within the landscape of physical systems provided by cQED setups, the objects known as geometric phases (GPs) have lately played an important role, mainly for implementing measurements and other quantum operations and therefore allowing {GPs} 
to be harnessed for various applications, such as geometric gates. 

The idea that the phase acquired by the state of a quantum system can be decomposed into a dynamical and a geometrical component originated with Berry's theoretical work, where it was constrained to the context of adiabatic, cyclic, unitary evolution~\cite{1JC}. Subsequently, the concept of GP has been extended to non-adiabatic cyclic, non-cyclic, and even non-unitary evolutions~\cite{2JC,3JC,4JC,5JC,6JC,7JC,8JC,9JC}. These generalizations consistently reduce to less comprehensive results as additional conditions are met. The GP has also been elucidated as a consequence of quantum kinematics, interpreted in terms of a parallel transport condition dependent solely on the geometry of the Hilbert space, from which its name is derived~\cite{10JC}. 

As evident in the extensive literature, GPs have evolved into not only a fruitful avenue for exploring fundamental aspects of quantum systems but also a topic of technological interest. For instance, due to its resilience to fluctuations in a coupled bath, the GP has been proposed as a significant resource for constructing phase gates~\cite{11JC,17JC, 18JC} for quantum information processing. Superconducting circuits have been extensively investigated as the physical system allowing this aim~\cite{kamleitner2011geometric, song2017continuous, abdumalikov2013experimental, leibfried2003experimental, xu2020experimental}. In a more fundamental approach, Berry phase has also been theoretically studied~\cite{38GP} and measured~\cite{40GP, leek2007observation} in several circuit architectures. The vacuum GP accumulated by a superconducting artificial atom that was interacting with a single mode of a microwave cavity was measured~\cite{41GP} as well, while the corrections on the GP introduced by transitions to higher exited levels of the transmon were examined in~\cite{42GP}.

When dealing with non-unitary dynamics leading to mixed states, the GP needs further generalization from its pure-state definition. A well defined proposal that applies under these conditions was presented in~\cite{8JC}. Thereafter, this definition has been applied to measure the corrections induced on the GP in non-unitary evolutions~\cite{19JC} and to explain the noise effects observed in the GP of a superconducting qubit~\cite{leek2007observation, 21JC}. Particularly, the GP of a two-level system under the influence of an external environment has been studied in a wide variety of scenarios~\cite{707713,032338}.
Even though the GP is not an immediate reflection of the dynamics and can therefore remain robust to the effect of the environment, it differs, in the general case, from that accumulated by the associated closed system, as the evolution is now affected by non-unitary effects such as decoherence and dissipation. 
Under suitable conditions, the non-unitary GP can be measured through interferometric (atomic interference)~\cite{26JC,27JC}, spin echo~\cite{leek2007observation}, and NRM~\cite{19JC, 28JC} experiments.

In a previous study~\cite{viotti2022geometric}, we thoroughly examined the GP accumulated 
by a two-level system (TLS) that was interacting with a single mode of the quantized electromagnetic field within a dissipative cavity, a physical setup known as a dissipative Jaynes--Cummings (JC) model. Addressing the scenario frequently encountered in semiconductor cavity quantum electrodynamics (QED)~\cite{36JC}, the interaction between the atom-mode system and its environment was characterized by the flow of photons through the cavity mirrors and the continuous, incoherent pumping spontaneously exciting the TLS.

In the present paper, we extend the work on the JC model to encompass the scenario of a nonlinear transmon coupled to a transmission line or resonator. Additionally, both the transmon and resonator are coupled to two semi-infinite waveguides serving as the surrounding environments. We will investigate the dynamics of the composite system, both in the qubit sector and in the two-excitation sector.
Restricting to the one-excitation sector will allow for direct comparison of the results previously obtained, which implies the comparison of two different architectures in which atom--cavity dynamics emerge. It is worth highlighting a major difference between both studies even at this early stage, which is the non-monotonic behavior encountered in the GP under certain environmental conditions. Thereafter, to further explore the richer nature of the transmon atom, we also examine the GP and its environmentally induced corrections when the two-excitation levels are involved in the dynamics. 

In the next section, we will introduce the Hamiltonian describing the non-harmonic transmon-field system under investigation and the coupling of the composite system to the environment. In this section, we will also present some insights about the dynamic evolution of the system and the definition of the geometric phase. In Section~\ref{sec_one-exc}, we will describe the one-excitation subspace dynamics and the correspondence with previous results. The two-excitation space and the role of charging energy and non-harmonicity is discussed in Section~\ref{two-excitations}. Section~\ref{conclu} summarizes our main conclusions.

\section{Transmon with Atomic--Kerr Interaction}

Due to their substantial dimensions, stemming from the necessity of maintaining low charging energy (via large capacitance), transmon qubits inherently lend themselves to capacitive coupling with microwave resonators. This coupling is reflected in the transmon Hamiltonian 
$
    \hat H = 4E_c (\hat n - n_g)^2 - E_J cos\hat\varphi
$
by the substitution of the classical voltage source $V_g$ with the resonator, a quantized gate voltage $n_g \rightarrow - \hat n_r$, representing the charge bias of the transmon due to the resonator. The Hamiltonian of the combined system \mbox{is therefore~\cite{12E}}

\begin{equation}
    \hat H = 4E_c (\hat n + n_r)^2 - E_J cos\hat\varphi - \sum_m \hbar \omega_m \hat a_m^\dagger \hat a_m,
\end{equation}
where $\hat n = \hat Q/2e$ is the charge number operator, and $\hat\varphi = (2\pi/\Phi_0)\hat\Phi$ (mod $2\pi$) is the phase operator, defined by the charge and phase operators, respectively, of the quantum circuit. The charging energy is $E_c = e^2/2C_\Sigma$, with $C_\Sigma = C_J + C_S$ being the sum of the junction’s capacitance $C_J$ and the shunt capacitance $C_S$. 
The operator $\hat n_r$ can be written as $\hat n_r = \sum_m \hat n_m$, with $\hat n_m = (C_g/C_m) \hat Q_m/2e$ being the contribution to the charge bias to the $m$th resonator mode. In this expression, $C_g$ is the capacitance of the gate and $C_m$ is the corresponding associated resonator mode capacitance. It is usual to consider $C_g \ll C_\Sigma, C_m$. When assuming that the transmon frequency is much closer to one of the resonator modes than all the other modes, it is possible to truncate the sum over $m$ to a single term. In this single-mode approximation for the resonator or cavity, the Hamiltonian reduces to a single oscillator of frequency denoted by $\omega_r$ coupled to a transmon. 

Expressed in terms of creation and annihilation operators, in the single-mode approximation, the Hamiltonian for the transmon--resonator cavity reduces to

\begin{equation}
    \hat H = \hbar \omega_r \hat a^\dagger \hat a + \hbar \omega_q \hat b^\dagger \hat b - \frac{E_c}{2} \hat b^\dagger\hat b^\dagger \hat b\hat b - \hbar g (\hat b^\dagger - \hat b) (\hat a^\dagger - \hat a),  
\end{equation}
where the term $\hat n_r^2$ has been absorbed in the charging 
energy term of the resonator mode and therefore leads to a renormalization of the resonator frequency, which we omit for simplicity. The frequency of the mode of interest is $\omega_r$, and the coupling constant between the artificial atom and the resonator is given by the relation
$
    g = \omega_r\, \sfrac{C_g}{C_\Sigma} \left(\sfrac{E_J}{2E_c}\right)^{\sfrac{1}{4}} \,(\sfrac{\pi Z_r}{R_K})^{\sfrac{1}{2}}, 
$
\mbox{where $Z_r$ is} the characteristic impedance of the resonator mode and $R_K = h/2e^2$ is the resistance quantum. The above Hamiltonian can be simplified further in the experimentally relevant situation where the coupling constant is much smaller than the system frequencies, $\vert g\vert \ll \omega_r, \omega_q$. After rotating-wave approximation, the Hamiltonian reads 

\begin{equation}
    \hat H_K \approx \hbar \omega_r \hat a^\dagger \hat a + \hbar \omega_q \hat b^\dagger \hat b - \frac{E_c}{2} \hat b^\dagger\hat b^\dagger \hat b\hat b + \hbar g (\hat b^\dagger \hat a + \hat b \hat a^\dagger).   \label{Hkerr}
\end{equation}
This is the Hamiltonian we shall consider to study the interaction between the artificial atom (in the transmon regime $E_J/E_c \gg 1$) and the electromagnetic field mode of the resonator. In this Hamiltonian, the term proportional to $E_c$ is the so-called Kerr-like interaction term, or non-harmonic term. 

\subsection{Coupling to the Environment}

Hitherto, our focus has been on quantum systems completely isolated from the surrounding environment. Nevertheless, a comprehensive portrayal of quantum circuits necessitates consideration of the manner in which these systems engage with their environment, encompassing both measurement apparatus and control circuitry. Indeed, the environment assumes a dual function in quantum technology; portraying quantum systems as entirely isolated is not only impossible due to inevitable coupling with undesirable environmental degrees of freedom but also renders a perfectly isolated system impractical for manipulation. Such a system would lack utility since we would be devoid of the means to control or observe it.
Given these considerations, in this section, we investigate our quantum system coupled to external semi-infinite {transmission lines that represent the measurement and control mechanisms while constituting the primary environment leading to photon loss and spontaneous decay of the transmon as their main environmental effect, respectively. We also consider the existence of a flux line for tunability of the transmon that leads to dephasing due to flux noise.}

\begin{figure}[ht!]
    \includegraphics[width = \linewidth]{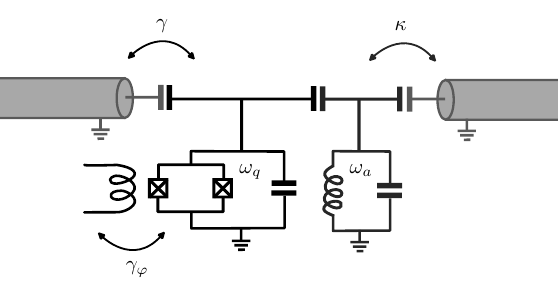}
    \caption{Transmon capacitively coupled to the resonator and both of them capacitively coupled to transmission lines. {The coefficient $\kappa$ is associated with the coupling between the resonator and the {readout} transmission line and represents the photon decay rate. In addition, $\gamma$ is related to the qubit-environment coupling strength introduced by the control mechanism and represents the relaxation rate. Finally, $\gamma_\varphi$ is the pure dephasing rate emerging due to the flux noise taking place in the flux line that allows for tunability of the transmon frequency}. \label{fig:Esquema}}  
\end{figure}

To study the open system comprising the transmon--resonator and transmission lines (reservoir), we will employ the conventional formalism provided by Lindblad master equations. In this context, we can model the transmission lines as a set of harmonic oscillators, similar to the approach taken in quantum Brownian motion models, which are paradigmatic examples of open quantum systems. Assuming, therefore, that the transmon and the resonator are coupled to independent baths (this system is illustrated in Figure~\ref{fig:Esquema}), the master equation for the composite system can be expressed as follows:

\begin{equation}
    \dot \rho = - i [\hat H_K, \rho] + \kappa {\cal D}[\hat a] \rho + \gamma {\cal  D}[\hat b] \rho 
+ \gamma_\varphi {\cal  D}[\hat b^\dagger \hat b^\dagger] \rho, \label{mastereq}
\end{equation}
where $\rho$ is the density matrix of the composite system (transmon--resonator), and $\hat H_K$ is the Hamiltonian of Equation~(\ref{Hkerr}). While this equation may suggest that dissipative processes independently impact the transmon and the resonator, the entanglement introduced by $\hat H_K$ implies that events such as the loss of a resonator photon can result in qubit relaxation. In this master equation, the coefficient $\kappa$ is the photon decay rate; $\gamma$ represents the relaxation rate of the artificial atom, which is related to the qubit-environment coupling strength evaluated at the qubit frequency; and $\gamma_\varphi$ is the pure dephasing rate that superconducting quantum circuits can also suffer, caused, for example, by fluctuations of parameters controlling their transition frequency and by dispersive coupling to other degrees of freedom in their environment. The symbol ${\cal D}[\hat {\cal O}] \rho$ in Equation~(\ref{mastereq}) represents the dissipator 

\begin{equation}
    {\cal D}[\hat {\cal O}] \rho = {\cal O} \rho {\cal O}^\dagger - \frac{1}{2} \left\{{\cal O}^\dagger {\cal O}; \rho \right\}, 
\end{equation}
where $ \left\{ . \, ; \, . \right\},$ is the anticommutator.

We will work on the basis ${\cal B} = \{|m~ n\rangle\}$, where $m$ refers to the $m$th energy eigenstate of the transmon and $n$ is the $n$th Fock state of the mode field in the resonator.

\subsection{Dynamic Evolution of the Open System}\label{sec:Th_dyn}

To address the dynamics of the transmon--field composite system, we numerically solve the master Equation~(\ref{mastereq}), constrained to the subspace with two or less excitations, \mbox{so that the} base $\mathcal{B}$
 reduces to $\{|00\rangle,|10\rangle,|01\rangle,|20\rangle,|11\rangle,|02\rangle\}$.
At a given instant $t$, the state of the system is described by a $6\times 6$ {density matrix} 

\begin{align}
\rho(t)= \left[
    \begin{array}{c|cc|ccc}
        \rho_{00} & \textcolor{gray}{\rho_{01}} & \textcolor{gray}{\rho_{02}} & \textcolor{gray}{\rho_{03}} & \textcolor{gray}{\rho_{04}} & \textcolor{gray}{\rho_{05}} \\ \hline
        \textcolor{gray}{\rho_{10}} & \rho_{11} & \rho_{12} & \textcolor{gray}{\rho_{13}} & \textcolor{gray}{\rho_{14}} & \textcolor{gray}{\rho_{15}} \\
        \textcolor{gray}{\rho_{20}} & \rho_{21} & \rho_{22} & \textcolor{gray}{\rho_{23}} & \textcolor{gray}{\rho_{24}} & \textcolor{gray}{\rho_{25}} \\ \hline
        \textcolor{gray}{\rho_{30}} & \textcolor{gray}{\rho_{31}} & \textcolor{gray}{\rho_{32}} & \rho_{33} & \rho_{34} & \rho_{35} \\
        \textcolor{gray}{\rho_{40}} & \textcolor{gray}{\rho_{41}} & \textcolor{gray}{\rho_{42}} & \rho_{43} & \rho_{44} & \rho_{45} \\
        \textcolor{gray}{\rho_{50}} & \textcolor{gray}{\rho_{51}} & \textcolor{gray}{\rho_{52}} & \rho_{53} & \rho_{54} & \rho_{55} \\
    \end{array}
    \right],
    \label{Eq:matrizdensidad}
\end{align}

which can be decomposed into blocks as in Equation~(\ref{Eq:matrizdensidad}). Elements belonging to different blocks satisfy differential equations that are decoupled so, by taking an initial condition with vanishing $\rho_{ij}$ elements for the off-diagonal blocks, the state of the system remains block diagonal along the whole evolution.

In this way, when the system is prepared in a one-excitation state $|\Psi(0)\rangle\in \{|10\rangle, |01\rangle\}$, the evolution remains restricted to the subspace form by the 2 {$\times$} 
 2 block and $\rho_{00}$. In that case, the dynamics renders independent on the value of the capacitance energy $E_c$ generating the anharmonicity. On the other hand, by setting the initial state as a pure state within the 3 $\times$ 3 block, all the matrix elements in the block-diagonal subspace are involved in \mbox{the evolution.}
\subsection{Geometric Phase in the Open System}

As already noted in the introductory Section~\ref{sec:intro}, a generalized definition of a GP that is suitable to be computed for a mixed state under non-unitary evolution was proposed in~\cite{7JC}. It reads

\begin{align}
        \phi_g[{\rho}] = \arg\;\bigg\lbrace\sum_{k=1}^N\,\sqrt{\epsilon_k(0)\epsilon_k(T)}\;\langle\psi_k(0)|\psi_k(T)\rangle \label{Eq:Tong_general}\\ \nonumber\times\, e^{-\int_0^T\,dt\,\langle\psi_k(t)|\Dot{\psi}_k(t)\rangle}\bigg\rbrace,
\end{align}
where $\psi_k(t)$ are the instantaneous eigenvectors of the density matrix, and $\epsilon_k(t)$ are the corresponding eigenvalues. This formula provides a well defined GP that, although defined for non-degenerate but otherwise general mixed states, when computed over pure states under unitary evolution, {reduces to the expression}
 {
 \begin{equation}
    \Phi_g(t)= \arg{\langle \Psi(0)|\Psi(t)\rangle} - {\rm Im} \int_0^t dt'~ \langle \Psi(t')|\dot{\Psi}(t')\rangle ,
\end{equation}}

\noindent
defined over the most general unitary evolution of a pure state $|\Psi(t)\rangle$~\cite{3JC, 37JC}.
It is also manifestly gauge invariant and therefore depends solely on the path traced by the state in the ray space.
 When dealing with pure initial states, $\epsilon_k(0) = 1$ for the specific $k$ labeling the initial state and vanishes for all $k'\neq k$. Therefore, Equation~(\ref{Eq:Tong_general}) reduces to a simpler form
 
\begin{equation}
    \Phi_g(t)= \arg{\langle \psi_+(0)|\psi_+(t)\rangle} - {\rm Im}\int_0^t dt'~ \langle \psi_+(t')|\dot{\psi}_+(t')\rangle 
    \label{Eq:Tong_2}
\end{equation}
where $|\psi_+(t)\rangle$ is the eigenvector of $\rho(t)$ that coincides, at $t=0$, with the initial state. This is the eigenvector such that $\epsilon_+(0) = 1$.
Equation~(\ref{Eq:Tong_2}) has the exact same functional form of the GP defined over unitary evolutions, for which the only difference is that the pure state involved is the eigenstate $|\psi_+(t)\rangle$ of the density matrix and not the state of the system itself{, which is now a mixed state}. 

We will restrict our analysis to pure initial states, so that the usual analysis applied to pure-state GPs can be immediately extrapolated by observing the behavior of the density matrix
eigenstate $|\psi_+(t)\rangle$.

\section{One-Excitation State: Dissipative Jaynes--Cummings Model}\label{sec_one-exc}
It will be useful to start by exploring some aspects of the better-known subspace accessed when preparing the system in a state with only one excitation. In this case, as stated in Section~\ref{sec:Th_dyn}, the state of the system remains constrained to the $2\times 2$ block spanned by $\{|10\rangle, |01\rangle\}$, with the only exception of showing population exchange to $\rho_{00}$. 

\begin{figure}[ht!] 
\centering
    \includegraphics[width=.85\linewidth]{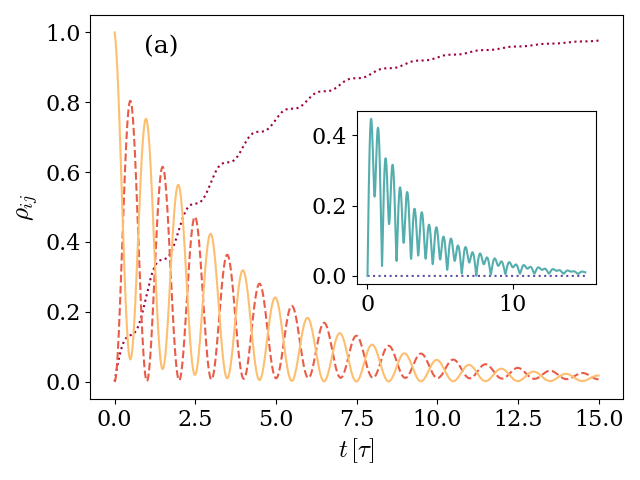}
    \includegraphics[width=.85\linewidth]{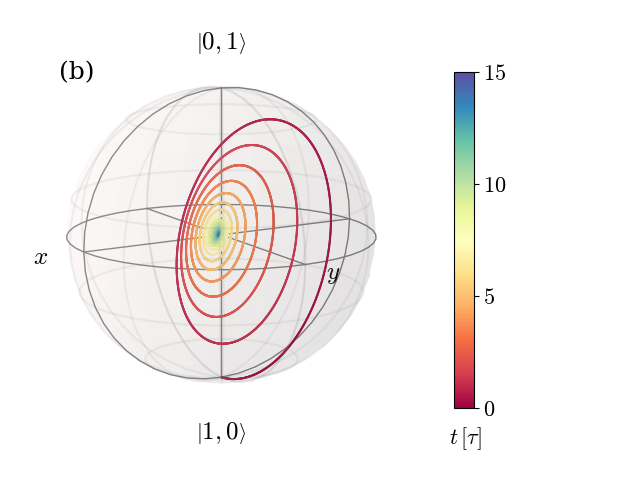}
    \caption{Dynamics of a system prepared in an initial $|\Psi(0)\rangle=|01\rangle$ and characterized by a detuning  {$\Delta = 0.017\, \omega_q$} and an atom-field coupling {$g=0.028\,\omega_q$}. The environment is described by a photon loss rate {$\kappa=0.005\,\omega_q$} and negligible atom relaxation $\gamma$ and dephasing $\gamma_\varphi$ rates.
    Panel (\textbf{a}) displays the density matrix elements evolution, and panel (\textbf{b}) shows the Bloch sphere representation of the $2\times 2$ block.
    In panel (\textbf{a}), the main plot displays the evolution of populations $\rho_{00}, \rho_{11}$, and $\rho_{22}$ with dotted purple, dashed orange, and solid yellow lines, whereas the inset displays the absolute value $|\rho_{12}|$. \label{fig:evolution_2x2}}  
\end{figure}

On general grounds, the populations $\rho_{11}$ and $\rho_{22}$ of energy levels with one excitation oscillate while decaying, whereas the population of the vacuum state $\rho_{00}$ grows. Depending on the parameters, the asymptotic state can be a pure $|00\rangle$ state or a mixed state with non-vanishing but suppressed $\rho_{11}$ and $\rho_{22}$ populations.
The only non-zero coherences $\rho_{12/21}(t)$ increase in absolute value up to a maximum value and then  vanish asymptotically. 
Figure~\ref{fig:evolution_2x2} shows the explicit situation in which the state is prepared in a state $|\Psi(0)\rangle = |01\rangle$ with one field excitation. 

It is useful to notice already at this point that the relation between the parameters defining the unitary evolution and the parameters defining the environmental effects results in a dynamic that exhibits specific characteristics. This is noticeable in Figure~\ref{fig:evolution_2x2}, where the absolute value of the only non-vanishing coherences reaches a minimum $|\rho_{12}|\sim 0$ at $t\sim 7.5\,\tau$, where it is smaller than the asymptotic value. This fact will be shown in different ways when observing the evolution of the eigenstate $|\psi_+(t)\rangle$ and the GP.
{In this archetypal example, we use $\omega_q = 2\pi\times 6$ GHz, $g = 2\pi\times 166.85$ MHz, and $\kappa = 2\pi\times 30$ MHz, which constitute up-to-date typical values~\cite{ahmad2023investigating, lledo2023cloaking, swiadek2023enhancing, cohen2207reminiscence}, while we consider a non-dispersive detuning $\Delta = \omega_q - \omega_r$ originating from a resonator frequency $\omega_r = 2\pi\times 5.97$ GHz.
In what follows, we will keep the typical values for the artificial atom frequency $\omega_q$ and the atom-mode coupling $g$ while inspecting the dynamics arising in different conditions defined by different parameter values. In this sense, the atom-mode detuning $\Delta$ will be modified from closer-to-resonance values $\Delta \sim \mathcal{O}(10)$~MHz to values within the dispersive regime $\Delta \sim \mathcal{O}(1)$ GHz. The artificial atom decay rate will also be increased to $\gamma =  2\pi\times 30$~MHz.} On the other hand, along the entire work, time is measured in units of $\tau = 2\pi/\Omega$, with the JC--Rabi frequency $\Omega = \sqrt{\Delta^2 + 4 g^2}$.

\begin{widetext}

    \begin{figure}[ht!]
    \centering
    \includegraphics[height=5.5cm]{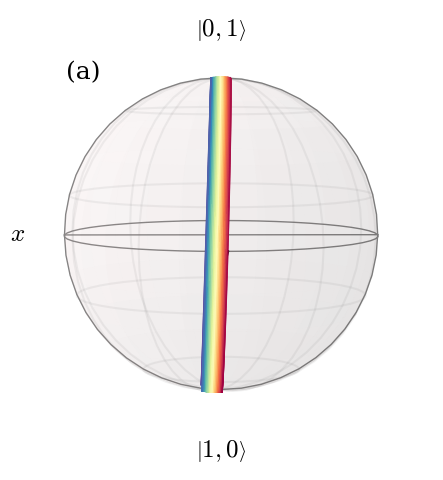}
    \includegraphics[height=5.5cm]{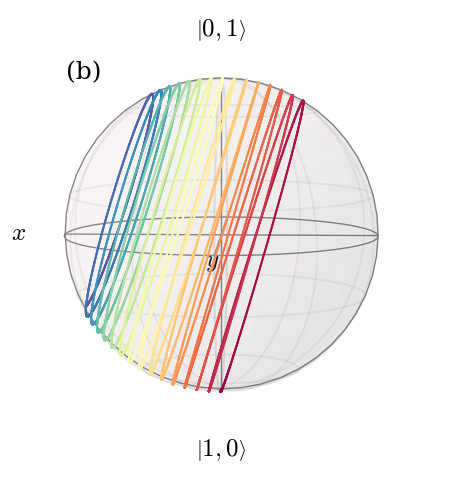}
    \includegraphics[height=5.5cm]{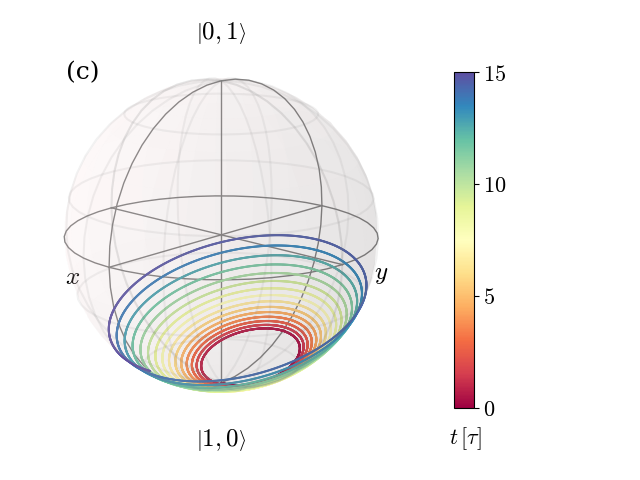}
    \caption{Trajectory displayed on the Bloch sphere by the eigenstate $|\psi_+(t)\rangle$ of the density matrix for the non-unitary evolution of a system prepared in an initial state $|\Psi(0)\rangle = |01\rangle$ for different ratios of the detuning $\Delta$ to the artificial atom frequency $\omega_q$. Panels (\textbf{a}--\textbf{c}) show the cases with {$\Delta = 0.0017\,\omega_q$}, {$\Delta = 0.017\,\omega_q$}, and {$\Delta = 0.17\,\omega_q$}, respectively. The environment and remaining features of the system are the same in all three panels. The environment is characterized by a photon loss rate {$\kappa=0.005\, \omega_q$} and negligible atom decay and dephasing rates, and the atom-field coupling considered satisfies the relation {$g=0.028\, \omega_q$.} \label{fig:Bloch_x3}}
    \end{figure}  
\end{widetext}

The eigenvector $|\psi_+(t)\rangle$ of the density matrix that is involved in the expression for the geometrical phase belongs, in this case, to the $2 \times 2$ block and can thus be observed on the Bloch sphere. In order to inspect the dependence of the dynamics with the detuning $\Delta$, Figure~\ref{fig:Bloch_x3} shows the evolution of $|\psi_+(t)\rangle$ on the Bloch sphere for a system that is prepared in a state $|\Psi(0)\rangle = |01\rangle$ with one field excitation for three different relations of the detuning $\Delta$ {to the frequency $\omega_q$ associated with the artificial atom}. The remaining conditions are taken to be equal in all three cases.
Panel (a) exhibits the case with {$\Delta = 0.0017\omega_q$}, panel (b) displays the case where {$\Delta = 0.017\omega_q$}, and panel (c) shows the case with {$\Delta = 0.17\omega_q$}. In all plots, the time is given by the color according to the color bar on the very right of the figure.

Figure~\ref{fig:Bloch_x3} shows that the path described by $|\psi_+(t)\rangle$ on the Bloch sphere is in all three cases a spiral that starts in the south pole of the sphere. The axis along which the spiral winds and moves differs in all three cases.
Under the conditions in panel (a) of Figure~\ref{fig:Bloch_x3}, in which the system is closer to resonance, the spiral axis is almost the $x$-axis and the curve moves little along it. Thus, the state traces a path that slightly deviates from vertical rings. When increasing the detuning, the axis of the spiral tilts and the turns separate from each other, as visible in panel (b). By further increasing the detuning, the axis of the spiral gets closer and closer to the z-axis and the initial turns get again closer while spreading for longer times.

\begin{figure}[ht!] 
    \centering
    \includegraphics[width=.8\linewidth]{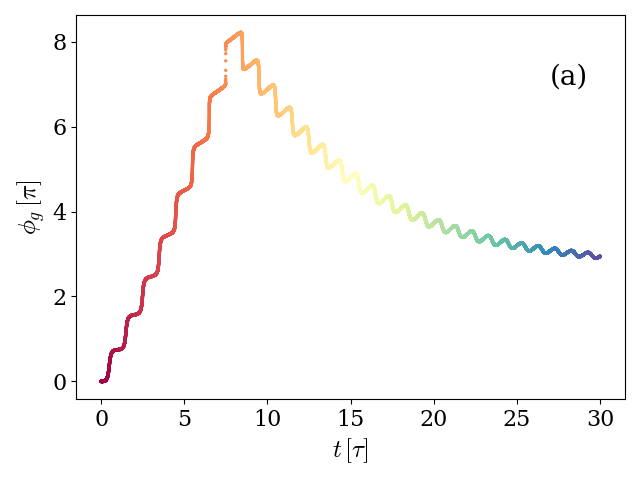}
    \includegraphics[width=.85\linewidth]{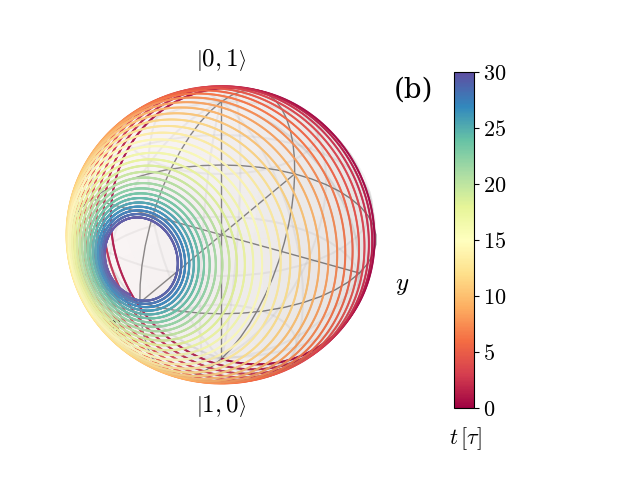}
 \caption{(\textbf{a}) Geometric phase accumulated over time by an initial state $|\Psi(0)\rangle=|01\rangle$ and (\textbf{b}) Bloch sphere depicting the path traced by the density matrix eigenvalue $|\Psi_+(t)\rangle$ for a system with detuning {$\Delta = 0.017 \,\omega_q$} and atom-field coupling {$g = 0.028\,\omega_q$}. The environment is characterized by a photon loss rate {$\kappa = 0.005\,\omega_q$} for the artificial atom and negligible atom decay $\gamma$ and dephasing $\gamma_\varphi$ rates. The color depicts, in both panels, the time instant as indicated by the color bar on the {right}.\label{fig:GP_Bloch_01}}  
\end{figure}

In some cases, as displayed in panels (a) and (b) of Figure~\ref{fig:Bloch_x3}, this behavior implies the exploration of different hemispheres of the sphere, in which the winding starts on one side of a certain (different in each case) great circle and crosses to the other side at some point. When this happens, the GP accumulated changes sign. In order to see this, Figure~\ref{fig:GP_Bloch_01} shows both the GP accumulated as a function of time (a), and the corresponding path traced by the $|\psi_+(t)\rangle$ eigenstate on the Bloch sphere (b), for the characteristic example depicted in Figures \ref{fig:evolution_2x2} and \ref{fig:Bloch_x3}b. 

As expected, the accumulated GP has a sign while the state is winding the sphere on one hemisphere, and it changes sign as soon as the winding takes place on the other hemisphere. It can be seen from Figure~\ref{fig:evolution_2x2} that this occurs at the specific time instant in which the coherences reach their minimum value. Therefore, we can classify the dynamics of the system initialized with one excitation as two kinds. One kind of evolution is that in which the coherences reach a minimum lower than the asymptotic value, the winding of the eigenstate $|\psi_+(t)\rangle$ changes hemisphere, and the GP changes sign, whereas the other is that in which none of these things happen.

\begin{figure}[ht!]
\centering
    \includegraphics[width=.9\linewidth]{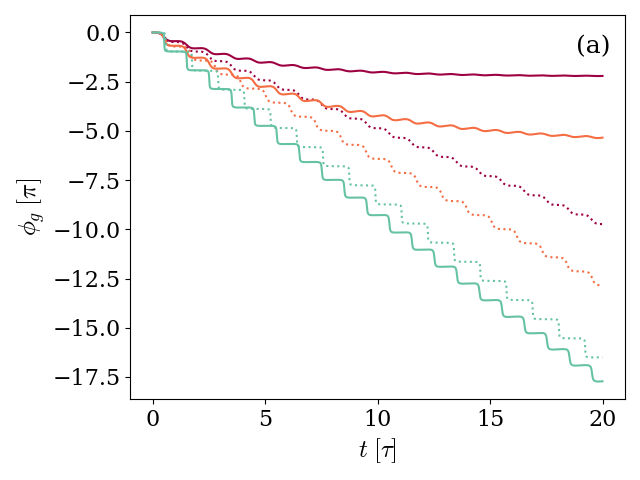}
    \includegraphics[width=.9\linewidth]{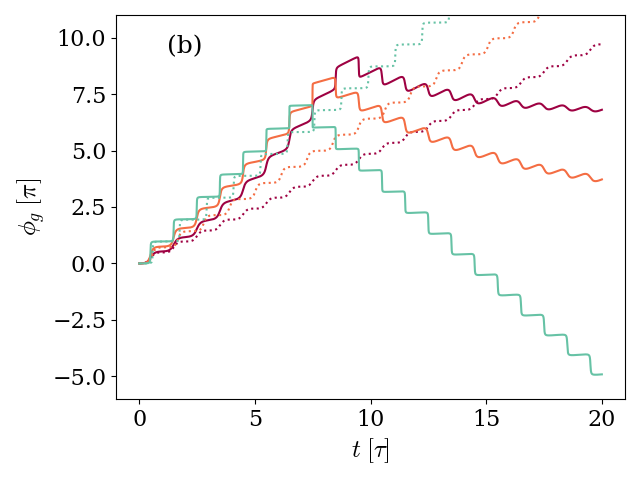}
    \caption{Geometric phase accumulated by systems with different values of detuning $\Delta$, but otherwise equal. In panel (\textbf{a}) the system is prepared in an initial state $|\Psi(0)\rangle=|10\rangle$, whereas panel (\textbf{b}) shows the case with $|\Psi(0)\rangle=|01\rangle$. The detuning values {$\Delta/\omega_q = 0.0017,\, 0.017$, and 0.34} are depicted by the {light blue, red, and purple} 
 solid lines, respectively. The atom-field coupling satisfies {$g = 0.028\,\omega_q$}, and the environment is characterized by photon loss rate {$\kappa = 0.005\,\omega_q$} and negligible \mbox{atom-decay $\gamma$ and} dephasing $\gamma_\varphi$ rates. The unitary results are included as dotted lines for reference, following the same $\Delta/\omega_q$-to-color code.\label{fig:GP_kappa}}  
\end{figure}

\begin{figure}[ht!]
   \includegraphics[width=.9\linewidth]{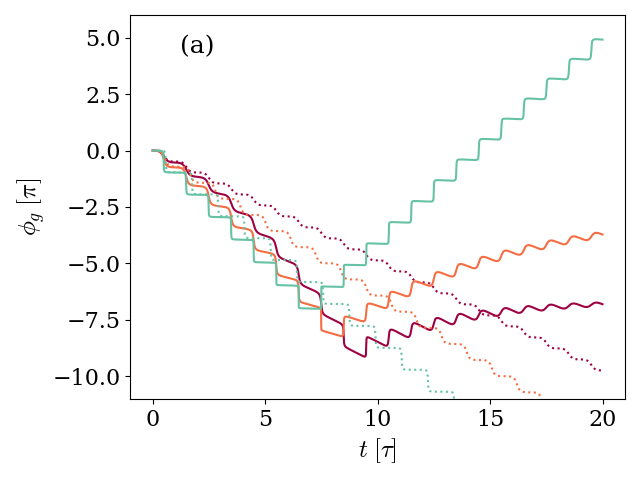}
   \includegraphics[width=.9\linewidth]{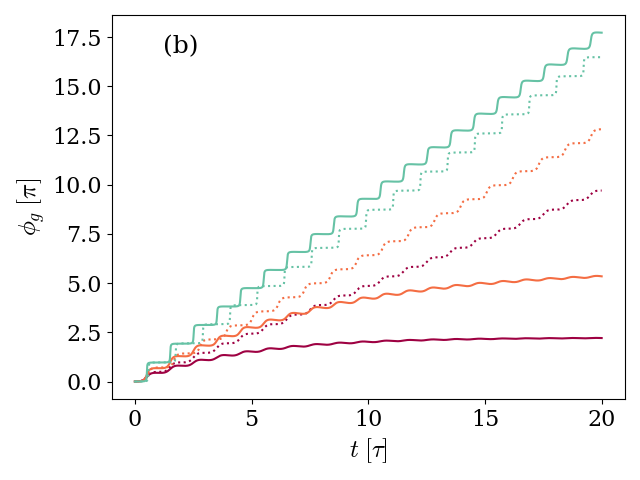}
 \caption{Geometric phase accumulated by systems with different values of detuning $\Delta$, but otherwise equal. In panel (\textbf{a}) the system is prepared in an initial state $|\Psi(0)\rangle=|10\rangle$, whereas panel (\textbf{b}) shows the case with $|\Psi(0)\rangle=|01\rangle$. The detuning values {$\Delta/\omega_q = 0.0017,\, 0.017$, and 0.34} are depicted by the {light blue, red, and purple} 
 solid lines, respectively. The atom-field coupling satisfies {$g = 0.028\,\omega_q$}, and the environment is characterized by a decay rate {$\gamma = 0.005\,\omega_q$} and negligible photon loss $\kappa$ and dephasing  $\gamma_\varphi$ rates. The unitary results are included as dotted lines for reference, following the same $\Delta/\omega_q$-to-color code. \label{fig:GP_gamma}}
\end{figure}

To better explore under which physical circumstances the GP accumulation is non-monotone, we study the dependence in three characteristics of the system and environment. These are: the initial state of the system, which we take to be either $|10\rangle$ or $|01\rangle$; the main source of decoherence and dissipative effects, which we consider to be either the photon loss $\kappa$ or the atom spontaneous decay $\gamma$; and the detuning $\Delta$. 
The results of this examination are displayed in Figures \ref{fig:GP_kappa}
and \ref{fig:GP_gamma}, which show the GP accumulated in time for different combinations of the parameters characterizing the system. In both figures, solid lines represent the GP accumulated by the dissipative system, and the unitary results are introduced as dotted lines for reference.

Figure~\ref{fig:GP_kappa} corresponds to the previously explored case in which the photon loss process is the main source of dissipation. 
Three different ratios of the detuning to the frequency associated with the artificial atom {$\Delta = 0.0017\,\omega_q$}, {$\Delta = 0.017\,\omega_q$}, and {$\Delta = 0.034\,\omega_q$} are displayed.  Panel (a) shows the GP accumulated by a system prepared in the first excited level of the transmon atom $|\Psi(0)\rangle = |10\rangle$, and, in panel (b), the initial state has a single field excitation $|\Psi(0)\rangle = |01\rangle$.

The two kinds of evolution, giving rise to monotonic or non-monotonic GPs, are clearly observed in Figure~\ref{fig:GP_kappa}. In panel (a), the GP accumulated by an initial $|10\rangle$ state is softer than the unitary result due to the environmental effects on the dynamics. When the state reaches the steady state and therefore stops moving on the ray space, the GP settles. On the other hand, in panel (b), the GP accumulated by an initial $|01\rangle$ is non-monotonic, with the change of direction found sooner for smaller $\Delta/\omega_q$ ratios.
Therefore, the dynamics leading to non-monotonic GPs are found when the main environmental effects are those affecting the initial excitation of the system. It is worth noticing that the results in Figure~\ref{fig:GP_kappa} are in full agreement with the results obtained in~\cite{viotti2022geometric}, in which the explored situation was that of a system prepared in  an initial $|10\rangle$ state and afterwards evolving in an imperfect cavity, corresponding to the case displayed in panel (a).

Likewise, Figure~\ref{fig:GP_gamma}, where the GP accumulated in time is shown for the same three ratios of the detuning to the frequency of the transmon {$\Delta = 0.0017\,\omega_q$}, {$\Delta = 0.017\,\omega_q$}, and {$\Delta = 0.034\,\omega_q$}, but the main environmental effect is the atom spontaneous decay. Once again, panel (a) shows the case in which the system is prepared in the first excited level of the transmon atom $|\Psi(0)\rangle = |10\rangle$, whereas in panel (b) the initial state has a single field excitation $|\Psi(0)\rangle = |01\rangle$.

Consistent with the statement that non-monotonic GPs are found when the main environmental effects are those affecting the initial excitation of the system, in Figure~\ref{fig:GP_gamma} \mbox{a change} in the GP direction is only observed in panel (a), in which the initial state is the first excited level of the transmon atom and vacuum field $|\Psi(0)\rangle = |10\rangle$.
As was also found in Figure~\ref{fig:GP_kappa}, the strongest the atom decay rate in relation to the frequencies associated with unitary evolution, the sooner the GP accumulation changes sign.

Comparing to previous results in~\cite{viotti2022geometric}, the absence of non-monotonic behavior found there can be explained as a combination of both the initial state in which the atom--photon system was prepared and the main environmental phenomena affecting it when the physical architecture is semiconductor cavities. In that case, the effect considered in Figure~\ref{fig:GP_gamma} was absent.

\begin{figure}[ht]
    \centering
    \includegraphics[width=.85\linewidth]{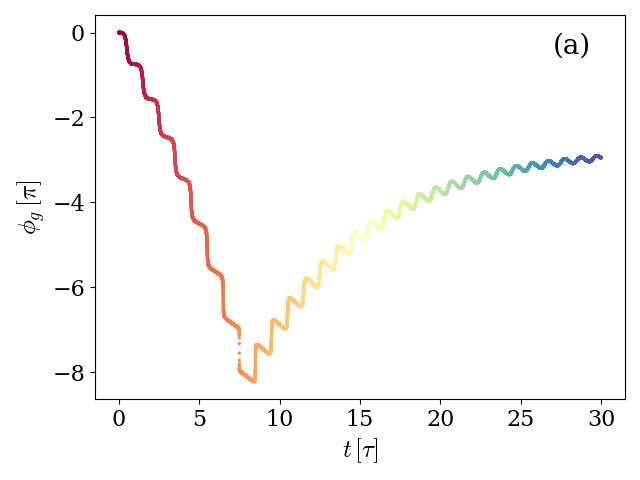}
    \includegraphics[width=.85\linewidth]{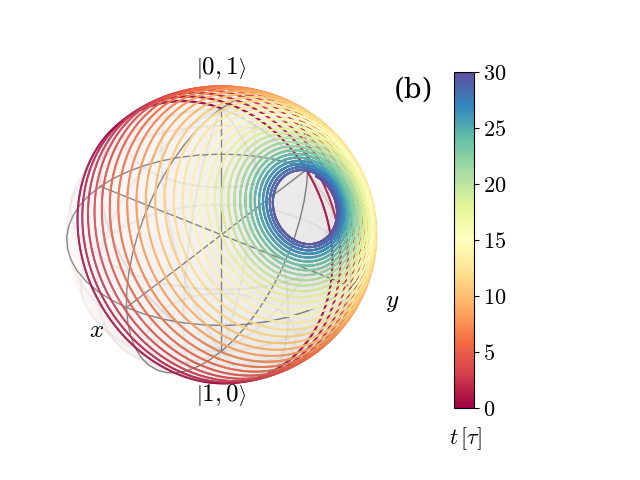}
    \caption{(\textbf{a}) Geometric phase accumulated over time by an initial state $|\Psi(0)\rangle=|10\rangle$ and (\textbf{b}) Bloch sphere depicting the path traced by the density matrix eigenvalue $|\psi_+(t)\rangle$ for a system with detuning {$\Delta = 0.017\,\omega_q$} and spin-field coupling {$g = 0.028\,\omega_q$}. The environment is characterized by a relaxation rate {$\gamma = 0.005\,\omega_q$} for the artificial atom and negligible photon loss $\kappa$ and dephasing $\gamma_\varphi$ rates. The color depicts, in both panels, the time instant, as indicated by the color bar on the {right}. \label{fig:Gp_Bloch_10}}  
\end{figure}

The evolution in Figure~\ref{fig:GP_gamma} can be re-observed, giving emphasis to the time instants by displaying it in the same manner as Figure~\ref{fig:GP_Bloch_01}. The analog plots compose Figure~\ref{fig:Gp_Bloch_10}, which thus shows the GP accumulated as a function of time (a) and the corresponding path traced by the $|\psi_+(t)\rangle$ eigenstate on the Bloch sphere (b) for a system prepared in  the first excited level of the transmon atom and vacuum field, with detuning-to-frequency rate {$\Delta =0.017\,\omega_q$}, and atom decay rate {$\gamma = 0.005\,\omega_q$}.

Once again, the change in the direction of the GP coincides in time with the moment in which the path traced by $|\psi_+(t)\rangle$ crosses from one side to the opposite of a great circle.

In order to re-state the description in terms of the behavior of the coherence, we go back to the initially considered case in which the initial state of the system has a single field excitation $|\Psi(t)\rangle = |01\rangle$. The analysis performed indicates there are two main situations in which the GP accumulated by this state will be monotonic: (a) if the relation between the environmental effects is such that the photon loss results are negligible in comparison with the atom decay rate, and (b) if, even though the main source of decoherence were the photon loss, the unitary parameters are strong enough to prevent the hemisphere crossing until the steady state is achieved.

\begin{figure}[ht!]
    \includegraphics[width=.9\linewidth]{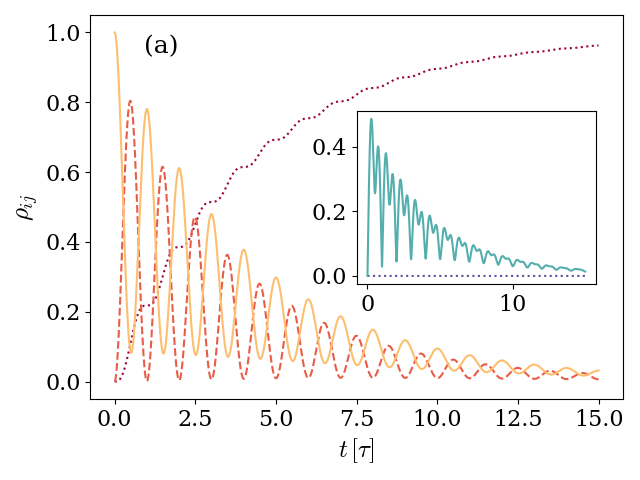}
    \includegraphics[width=.9\linewidth]{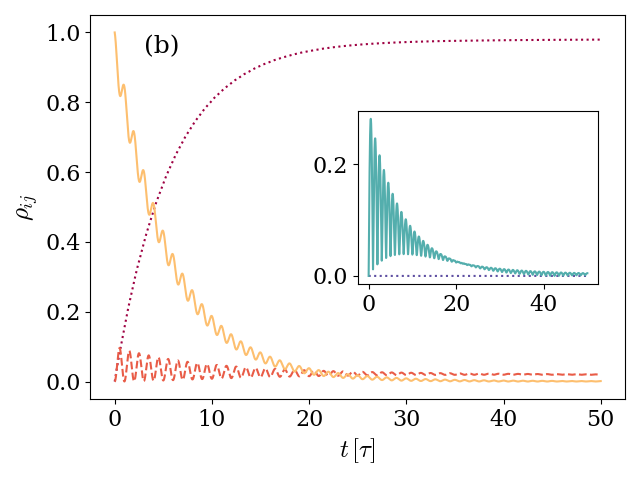}
    \caption{{Density} 
 matrix elements evolution in time for two different conditions under which the GP does not change monotony.
    In both cases, the main plot displays the evolution of populations $\rho_{00}, \rho_{11}$, and $\rho_{22}$ with dotted purple, dashed orange, and solid yellow lines, and the insert displays the absolute value $|\rho_{12}|$ (solid light-blue line) and the vanishing of the remaining coherences $|\rho_{i0}|$ (blue dotted line). Also in both panels, the system is prepared in an initial $|\Psi(0)\rangle=|01\rangle$ with atom-field coupling {$g = 0.028\,\omega_q$}.
    Panel (\textbf{a}) shows the case with {$\Delta = 0.0017\,\omega_q$}, {$\gamma=0.005\,\omega_q$}, and negligible photon loss $\kappa$ rate, whereas panel (\textbf{b}) addresses the case with {$\Delta = 0.017\,\omega_q$}, {$\kappa=0.005\,\omega_q$}, and negligible atom decay $\gamma$ rate. Dephasing is considered a subleading process on both plots. \label{fig:coherencias_2x2}}  
\end{figure}

Figure~\ref{fig:coherencias_2x2} shows the density matrix elements evolution for both these situations in panels (a) and (b), respectively.

It is immediately noticed that, in none of those cases, the absolute value of the coherences goes below its asymptotic value. In panel (a), where the photon loss is negligible in comparison with the atom decay rate, the excited populations $\rho_{11}$ and $\rho_{22}$ decay oscillating as the ground state gets populated. Different from what was observed in Figure~\ref{fig:evolution_2x2}, the absolute value of the non-zero coherences $|\rho_{12}|$ never  reached a minimum below its asymptotic value.
A similar behavior is observed in panel (b), where the increase in the detuning ratio $\Delta/\omega_q$ results in a relatively less strong environment that requires observation over longer timescales. As in panel (a), in this case the coherences absolute value is never below the asymptotic value.

\section{Two-Excitation Space: Role of Charging Energy and Non-Harmonicity} \label{two-excitations}
When turning to higher excited initial states, the evolution of the system takes place in the full six-dimensional Hilbert space described in Section~\ref{sec:Th_dyn}. As stated there, if the system is prepared in a state with a defined number of excitations, the originally vanishing coherences remain zero at all following times. 

On general grounds, the excited populations decay as lower-energy states get populated and, in the same way as described in Section~\ref{sec_one-exc}, the system evolves to an asymptotic state that can be either the pure ground state $|00\rangle$ or a mixed state with non-vanishing but suppressed populations on exited states. The non-zero coherences increase in absolute value up to a maximum value, to vanish asymptotically afterwards.

\begin{figure}[ht!]
   \includegraphics[width=.85\linewidth]{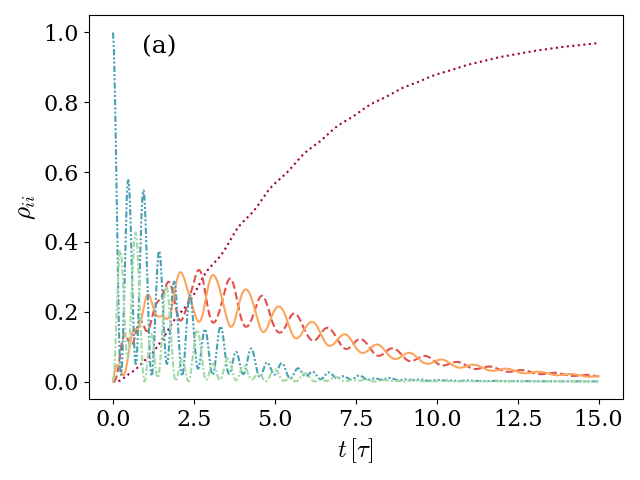}
   \includegraphics[width=.85\linewidth]{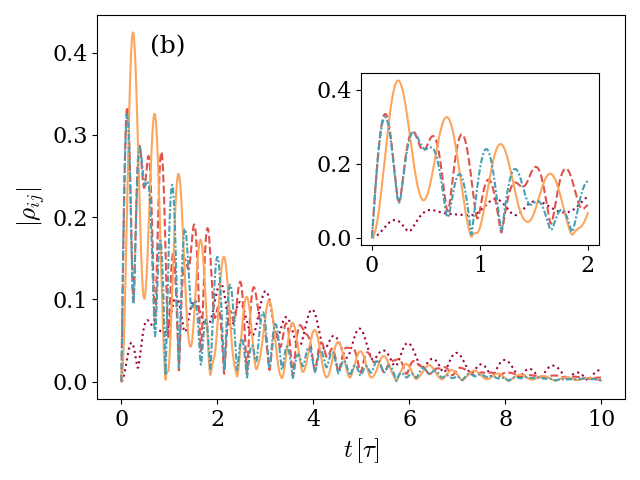}
 \caption{Dynamics of the system, depicted by the time evolution of the density matrix elements. Panel (\textbf{a}) shows the evolution of the populations $\rho_{ii}(t)$, with line stiles associated with matrix elements as follows: the dotted purple line shows $\rho_{00}$, the dashed red line shows $\rho_{11}$, the solid orange line shows $\rho_{22}$, the dot-dashed green line shows $\rho_{33}$, and the double dot-dashed light blue line shows $\rho_{44}$. \mbox{Panel (\textbf{b})} displays the evolution, in absolute value, of the non-zero coherences $\rho_{ij}\,,i\neq j$. In this panel, the dotted purple line, dashed red line, solid orange line, dot-dashed green line, and double dot-dashed light blue line correspond to the matrix elements $\rho_{12}, \rho_{34}, \rho_{35}$, and $\rho_{45}$. The considered system is prepared in an initial state $|\Psi(0)\rangle = |11\rangle$ in the first excited level of the artificial atom and one field excitation. It is further characterized by a detuning value {$\Delta = 0.0017\,\omega_q$}, atom-field coupling rate of {$g = 0.028\,\omega_q$}, and anharmonicity {$E_c = 0.035\,\omega_q$}. The environment is characterized by photon loss {$\kappa = 0.005\,\omega_q$} rate and negligible atom decay $\gamma$ and dephasing  $\gamma_\varphi$ rates. The unitary results are included as dotted lines for reference. \label{fig:rho_kappa_3x3}}
\end{figure}

In order to make the most simple generalization possible of the one-excitation case, we consider in this section a system that is prepared in a two-excitation state $|\Psi(0)\rangle = |11\rangle$, with excitations of a different nature. Figure~\ref{fig:rho_kappa_3x3} shows an specific example depicting the above described behavior.

It can be observed in panel (a) of Figure~\ref{fig:rho_kappa_3x3} that the population $\rho_{44}(t)$ associated with the $|11\rangle$ state decreases and shows non-harmonic oscillations. Along with this decrement, an immediate increment of the $\rho_{33}(t)$ and $\rho_{55}(t)$ populations, associated with the remaining two-excitation states $|20\rangle$ and $|02\rangle$, takes place. These are suppressed within a few Rabi periods. Due to the effect of the environment, the $\rho_{11}(t)$ and $\rho_{22}(t)$ elements associated with one-excitation states also show an increment and afterwards decrement while oscillate harmonically. The timescale of the increment--decrement of these elements is longer than the timescale shown by the two-excitation populations. Finally, the $\rho_{00}(t)$ population associated with the ground state monotonically increases along the whole evolution up to a steady value $\sim$$1$. 
The non-zero coherences are now four (and their corresponding complex conjugates). Panel (b) of Figure~\ref{fig:rho_kappa_3x3} shows the evolutions of these elements. In all cases, the absolute value of these elements show an initial increment and asymptotic vanishing. Noticeably, the timescale associated with the $\rho_{12}(t)$ coherence belonging to the one-excitation subspace is larger than the timescale in which the coherences $\rho_{34}, \rho_{35}$, \mbox{and $\rho_{45}$,} belonging to the two-excitation block of $\rho(t)$, evolve.

\begin{figure}[ht]
   \includegraphics[width=.85\linewidth]{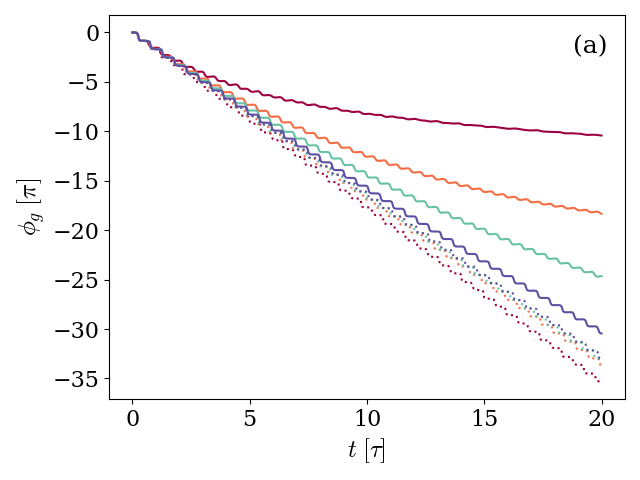}
   \includegraphics[width=.85\linewidth]{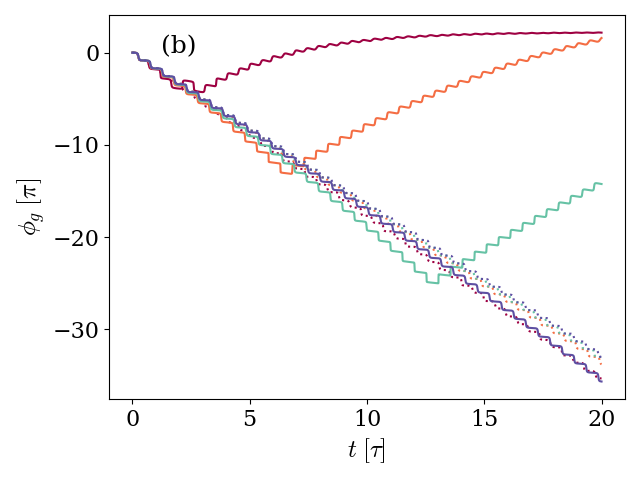}
 \caption{Geometric phase accumulated by a system prepared in an initial state with one excitation of each kind $|\Psi(0)\rangle=|11\rangle$ and characterized by different values of detuning $\Delta$, but otherwise equal. 
 The detuning values {$\Delta/\omega_q = 0.0017,\, 0.013,\,0.015$, and 0.017} are depicted by {the purple, red, light blue, and blue} 
 solid lines, respectively, the atom-field coupling satisfies {$g = 0.028\,\omega_q$}, and the anharmonicity is {$E_c = 0.035\,\omega_q$}.
 In panel (\textbf{a}), the environment is characterized by an atom decay rate {$\gamma = 0.005\,\omega_q$} and negligible photon loss $\kappa$ and dephasing  $\gamma_\varphi$ rates.
 On the other hand, in panel~(\textbf{b}), the environment is characterized by photon loss rate {$\kappa = 0.005\,\omega_q$} and negligible atom decay $\gamma$ and dephasing  $\gamma_\varphi$ ratios. The unitary results are included as dotted lines for reference, following the same $\Delta/\omega_q$-to-color code. \label{fig:GP_3x3}}
\end{figure}

In order to reproduce the analysis performed for evolutions constrained to the one and zero-excitation subspace, we turn now to the observation of the GP accumulated by the state of the system in time. With this purpose, Figure~\ref{fig:GP_3x3} shows the GP as a function of time for systems characterized by different ratios of the detuning $\Delta$ to the frequency $\omega_q$ associated with the transmon first transition, but otherwise equal. On each panel, the main environmental effect is of a different kind: in panel (a) the main environmental effect is the spontaneous decay of the transmon, whereas in panel (b) the main environmental effect is the photon loss.

In panel (a) of Figure~\ref{fig:GP_3x3}, where the initial $|11\rangle$ state evolves subjected to the spontaneous decay of the artificial atom, the GP accumulated along the evolution differs from the unitary GP, in that the non-unitary increment is slower and softer up to a point in which it is completely stopped by the meeting of a steady state that does not move on the ray space any more. The bigger the atom-decay rate relative to the unitary parameters, the strongest the effect of the environment on the dynamics, which is reflected in the same way by the GP.

In addition, panel (b) shows the evolution of the same system for the case in which the main environmental effect is the photon loss. Under these circumstances, the non-monotonic behavior of the GP already observed in Section~\ref{sec_one-exc} is recovered.
If the detuning is small enough, the environment has the effect of changing the sign in the GP accumulated, which afterwards tends to an asymptotic state as the system reaches a steady state that does not move in the ray space. 
However, the differences observed when comparing to Figure~\ref{fig:GP_kappa} are not only quantitative but also qualitative: the GP acquired for {$\Delta = 0.0017\,\omega_q$} not only changes direction but also does so more than once. Instead of observing a single minimum step, there are two.

With the anharmonicity between levels being one of the main differences between these state subspaces, we further inspect the effect of the ratio {$E_c/\omega_q$} with focus on this difference in the behaviors observed in Figures \ref{fig:GP_kappa}  and \ref{fig:GP_3x3}.

\subsection*{{Effect} 
 of the Anharmonicity}
\begin{figure}[ht!]
   \includegraphics[width=.85\linewidth]{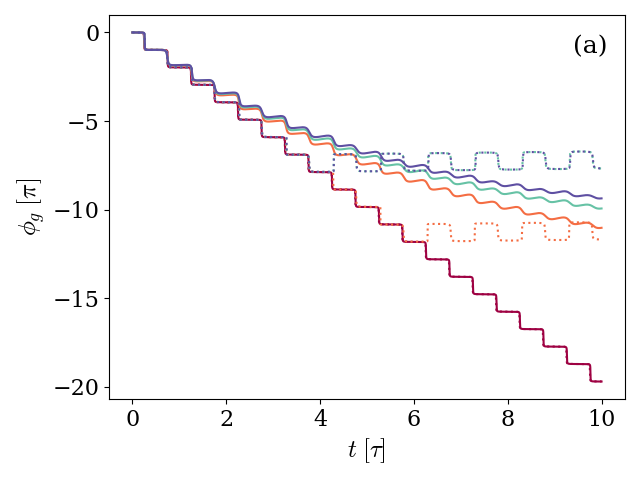}
   \includegraphics[width=.85\linewidth]{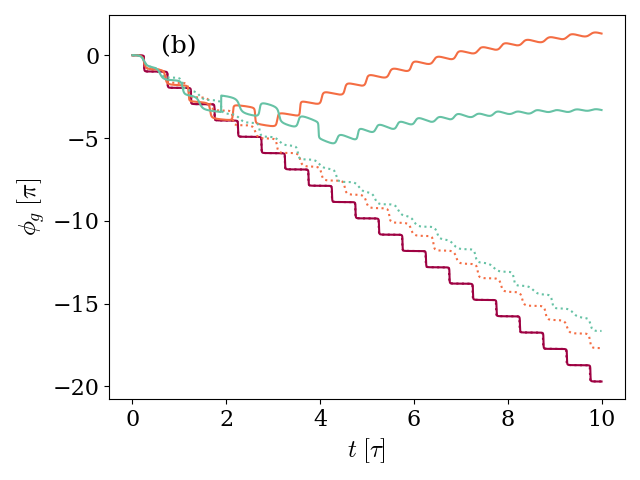}
 \caption{{Geometric} 
 phase accumulated by a system prepared in an initial state with one excitation of each kind $|\Psi(0)\rangle=|11\rangle$.
 Panel (\textbf{a}) reproduces (\textbf{b}), changing the values of the anhamonicity to {$E_c = 0.003\,\omega_q$}. 
 The considered detuning values are {$\Delta/\omega_q = 0.0017,\, 0.013,\,0.015$, and 0.017}, depicted by the purple, red, light blue, and blue solid lines, respectively, and the atom-field coupling satisfies {$g = 0.028\,\omega_q$}. The environment is characterized by a photon loss rate {$\kappa = 0.005\,\omega_q$} and negligible atom decay $\gamma$ and dephasing  $\gamma_\varphi$ rates. On the other hand, panel (\textbf{b}) shows the GP accumulated for systems with different values of the anharmonicity {$E_c/\,\omega_q = 0.003,\, 0.035$, and $0.067$}, depicted by the purple, red, and light blue solid lines, respectively. The system and environment are further characterized by a detuning {$\Delta = 0.0017 \,\omega_q$}, an atom-field coupling { $g = 0.028\,\omega_q$}, photon loss rate {$\kappa = 0.005\,\omega_q$}, and negligible atom decay $\gamma$ and dephasing $\gamma_\varphi$ rates. On both panels, the unitary results are included as dotted lines for reference, following the same (\textbf{a}) $\Delta/\omega_q$-to-color and (\textbf{b})~$E_c/\omega_q$-to-color~codes.  \label{fig:GP_3x3_Ec}} 
\end{figure}

In order to examine the effect of the anhamonicity $E_c$  on the GP accumulated by the non-unitary system, in panel (a) of Figure~\ref{fig:GP_3x3_Ec} we reproduce Figure~\ref{fig:GP_3x3}b for a different value of the anharmonicity $E_c$. In doing so, the behavior of both the unitary and non-unitary GP accumulated in time is qualitatively modified depending on the relation \mbox{between parameters.} 

With regard to the unitary results, displayed in the figure in dotted lines, the GP accumulated when decreasing the anharmonicity remains qualitatively equal to the previous situation for the smallest {$\Delta = 0.0017\omega_q$} ratio, as depicted by the purple dotted line. However, when increasing the detuning, the unitary GP reaches a regime in which it changes sign periodically, leading to a step-like oscillation around a fixed value. 

Regarding the GP accumulated in the presence of an environment, the non-monotonic behavior observed in Figure~\ref{fig:GP_3x3}b completely disappears, and the GP performs steps that get softer and smaller as time goes by up to the stationary regime in which the state stops moving in ray space, so no further GP is accumulated.

To further inspect this effect, panel (b) of Figure~\ref{fig:GP_3x3_Ec} shows the GP accumulated in time for a system that is prepared in the same $|\Psi(0)\rangle = |11\rangle$ state with fixed detuning ratio {$\Delta = 0.0017\omega_q$} and environment conditions described by {$\kappa = 0.005\,\omega_q$}, but different values {$E_c/\omega_q = 0.003,\, 0.035$, and $0.067$} of the anharmonicity $E_c$. With increasing anharmonicity, the non-monotonic behavior is recovered, and more changes are observed in the GP sign for \mbox{greater $E_c$ values.} 

As a whole, the GP accumulated by an initial state with one photon and the transmon in its first excited level shows the non-monotonic behavior only when the main effect of the environment is the photon loss, observed when both the anharmonicity splitting the transmon levels and the effect of the environment are big enough, leading to a system that is closer to a two-level system. Reducing the anharmonicity leads to a degeneracy in the artificial atom states that prevents the environmental non-monotonic behavior while introducing qualitative changes in the unitary GP.

\section{Discussion} \label{conclu}

In this paper we examined the dynamics beyond the two-level approximation of a transmon. Particularly, we studied the open dynamics of a nonlinear transmon coupled to a one-mode resonator and a transmission line. We have shown that the density matrix can be decomposed into blocks that satisfy differential equations that decoupled under particular initial conditions. Therefore, we were able to separately analyze one-excitation and two-excitation subspaces.
We further explored the geometric phase accumulated by the state in order to have a better insight into the richer nature of the transmon \mbox{artificial atom.}

In the case of a one-excitation state, we retrieved results of the dissipative Jaynes--Cummings model. However, as we contemplated complex Linblad equations (with three different noise channels), we can complete already existing results. For example, we found the existence of a non-monotonic behavior in the accumulated geometric phase due to a combination of both the initial state and the main leading noise ruling \mbox{the dynamics.}

In the case of the two-excitation state, we studied the dissipative dynamics of the system in a regime where $E_c$ cannot be neglected. The open dynamics are more complex, but we were able to present studies on the numerically accumulated geometric phase on a \mbox{3 $\times$ 3} subspace and study its behavior as a function of the anharmonicity rates. Again, the GP accumulated by an initial state with one photon and the transmon
in its first excited level shows the non-monotonic behavior only when the main effect of the environment is the photon loss. This can be better understood with the help of the complete analysis on the \mbox{2 $\times$ 2} subspace (where we can interpret results with the help of the Bloch sphere). 

It is important to remark that in the near-resonance case, the accumulated geometric phase seems to remain robust, as previous results stated. 
\\
\\
This research was funded by Agencia Nacional de Promoción Científica y Tecnológica 
(ANPCyT), Consejo Nacional de Investigaciones Científicas y Técnicas (CONICET), and Universidad de 728
Buenos Aires (UBA).

\noindent 

\bibliography{bibfile.bib}

\end{document}